\documentclass[iop,apj,tighten]{emulateapj}
\usepackage[para,online,flushleft]{threeparttable}
\usepackage{etoolbox}
\usepackage{tablefootnote, natbib}
\usepackage{aas_macros}

\usepackage{amsmath}
\BeforeBeginEnvironment{appendices}{\clearpage}
\usepackage{tabularx}

\newcommand\clearrow{\global\let\rowmac\relax}
\clearrow

\appto\TPTnoteSettings{\footnotesize}

\usepackage[breaklinks,colorlinks,citecolor=cyan,linkcolor=cyan]{hyperref} 

\usepackage[all]{hypcap} 
\usepackage{graphicx}
\usepackage{relsize}

\usepackage{ulem}

\shorttitle{A Black Hole in M31's most massive globular cluster}
\shortauthors{Pechetti et al.}

\begin{document}
\title{Detection of a $\sim$100,000 M$_\odot$ Black Hole in M31's most massive globular cluster: A tidally stripped nucleus}
\author{Renuka Pechetti\altaffilmark{1}, Anil Seth\altaffilmark{2}, Sebastian Kamann\altaffilmark{1}, Nelson Caldwell\altaffilmark{3}, Jay Strader\altaffilmark{4}, Mark den Brok\altaffilmark{5}, Nora Luetzgendorf\altaffilmark{6}, Nadine Neumayer\altaffilmark{7} \& Karina Voggel\altaffilmark{8} }

\affil{\begin{center}
\textit{
{\scriptsize$^1$Liverpool John Moores University, UK\\
\scriptsize$^2$Department of Physics and Astronomy, University of Utah, 115 South 1400 East, Salt Lake City, UT 84112, USA\\
\scriptsize$^3$Harvard-Smithsonian Center for Astrophysics\\
\scriptsize$^4$Michigan State University, USA\\}
\scriptsize$^5$Leibniz-Institut f\"ur Astrophysik Potsdam, An der Sternwarte 16, 14482 Potsdam, Germany\\
\scriptsize$^6$European Space Agency, STScI, USA\\
\scriptsize$^7$Max Planck Instit\"ut f\"ur Astronomie, Heidelberg, Germany\\
\scriptsize$^8$Observatoire astronomique de Strasbourg, USA\\
}\end{center}}

\begin{abstract}

We investigate the presence of a central black hole (BH) in B023-G078, M31's most massive globular cluster. We present high-resolution, adaptive-optics assisted, integral-field spectroscopic kinematics from Gemini/NIFS that shows a strong rotation ($\sim$20 km/s) and a  velocity dispersion rise towards the center (37 km/s).  We combine the kinematic data with a mass model based on a two-component fit to $HST$ ACS/HRC data of the cluster to estimate the mass of a putative BH. Our dynamical modeling suggests a $>$3$\sigma$ detection of a BH component of 9.1$^{+2.6}_{-2.8}\times$10$^4$ M$_\odot$ (1$\sigma$ uncertainties). The inferred stellar mass of the cluster is 6.22$^{+0.03}_{-0.05}\times$10$^6$ M$_\odot$, consistent with previous estimates, thus the BH makes up 1.5\% of its mass.  We examine whether the observed kinematics are caused by a collection of stellar mass BHs by modeling an extended dark mass as a Plummer profile. The upper limit on the size scale of the extended mass is 0.56 pc (95\% confidence), which does not rule out an extended mass.  There is compelling evidence that B023-G078 is the tidally stripped nucleus of a galaxy with a stellar mass~$>$10$^9$ M$_{\odot}$, including its high mass, two-component luminosity profile, color, metallicity gradient, and spread in metallicity. Given the emerging evidence that the central BH occupation fraction of~$>$10$^9$ M$_{\odot}$ galaxies is high, the most plausible interpretation of the kinematic data is that B023-G078 hosts a central BH. This makes it the strongest BH detection in a lower mass ($<$10$^7$ M$_{\odot}$) stripped nucleus, and one of the few dynamically detected intermediate-mass BHs. 
\end{abstract}

\maketitle 

\section{Introduction}
Intermediate-mass black holes (IMBHs) are hypothesized to exist in the mass range between stellar-mass black holes ($\lesssim 100$~M$_\odot$) and super-massive black holes (SMBHs; $\gtrsim$10$^5$~M$_\odot$).  Some models of SMBH formation rely on stellar or IMBH mass seeds or direct collapse of gas clouds, and thus the detection or lack of IMBHs can help us understand the SMBH formation \citep[e.g.][]{greene20}.

Studying IMBHs and the lowest mass SMBHs in galaxy centers can also help in extending and understanding the correlations that exist between galaxies and their black holes \citep[ e.g.][]{gebhardt2000,mcconnell_ma13,saglia16} to lower masses.  

Recently, BHs with masses 10$^5$-10$^{7}$~M$_\odot$ have been detected in lower-mass galaxies with masses 10$^9$--10$^{10}$~M$_\odot$ using both dynamical measurements \citep{denbrok15,nguyen18,nguyen19,davis20}, and measurements of AGN \citep[e.g.][]{reines13, chilingarian18, mezcua18}. SMBHs with masses~$>$10$^{6}$~M$_\odot$ have also been found at the centers of ultracompact dwarfs \citep[UCDs; e.g.][]{seth14,ahn17}; massive star clusters that appear to be the tidally stripped nuclear star clusters of galaxies \citep[e.g.][]{mieske13,pfeffer13,neumayer20}.  While so far, these BHs have only been found in the highest mass UCDs \citep{voggel18}, there are likely lower mass stripped nuclei and BHs hiding among galaxies' globular cluster (GC) systems \citep{voggel19}. These objects are among the most likely targets for detecting IMBHs.  
\begin{figure*}
\begin{center}
\includegraphics[trim=2cm 10cm 2cm 4cm, clip=true, width=0.8\linewidth]{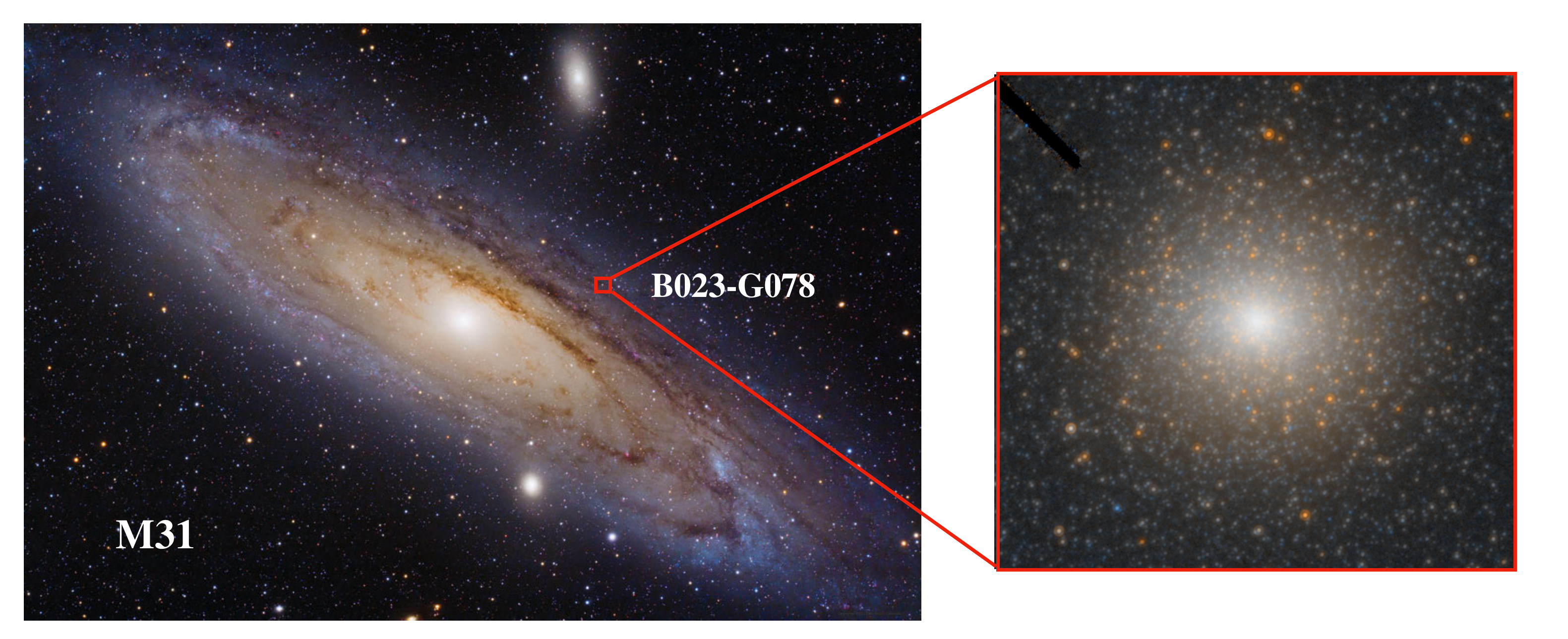}
\caption{Location and color image of B023-G78. The left panel shows a wide-field image of M31 (Image Credit: Iv\'{a}n \'{E}der, https://www.astroeder.com/), with the red box and inset showing the location and HST ACS/HRC image of B023-G78, which is $\sim$10$'' \times$10$''$. }

\label{fig:m31}
\end{center}
\end{figure*}

Although GCs are potential reservoirs for IMBHs, detecting these IMBHs remains challenging for several reasons. First, the gravitational sphere of influence of the IMBHs is small, which limits dynamical IMBH searches (that must resolve this radius) to within the Local Group. Second, dynamical evolution in GCs causes stellar-mass black holes (and more slowly, neutron stars) to mass segregate at the center of a cluster.  Collections of these stellar remnants can create a rise in the central velocity dispersion mimicking an IMBH \citep[e.g.,][]{zocchi19,baumgardt19}.  While many stellar-mass BHs will be lost due to interactions or natal kicks, a significant fraction of BHs can be retained at the center in some clusters
\citep[up to $\sim$2\% of the cluster's mass;][]{weatherford20}. Lastly, radial anisotropy can also contribute to creating an observed rise in the central velocity dispersion without the presence of an IMBH \citep{zocchi17}.

\begin{figure*}
\begin{center}

\includegraphics[width=0.9\linewidth]{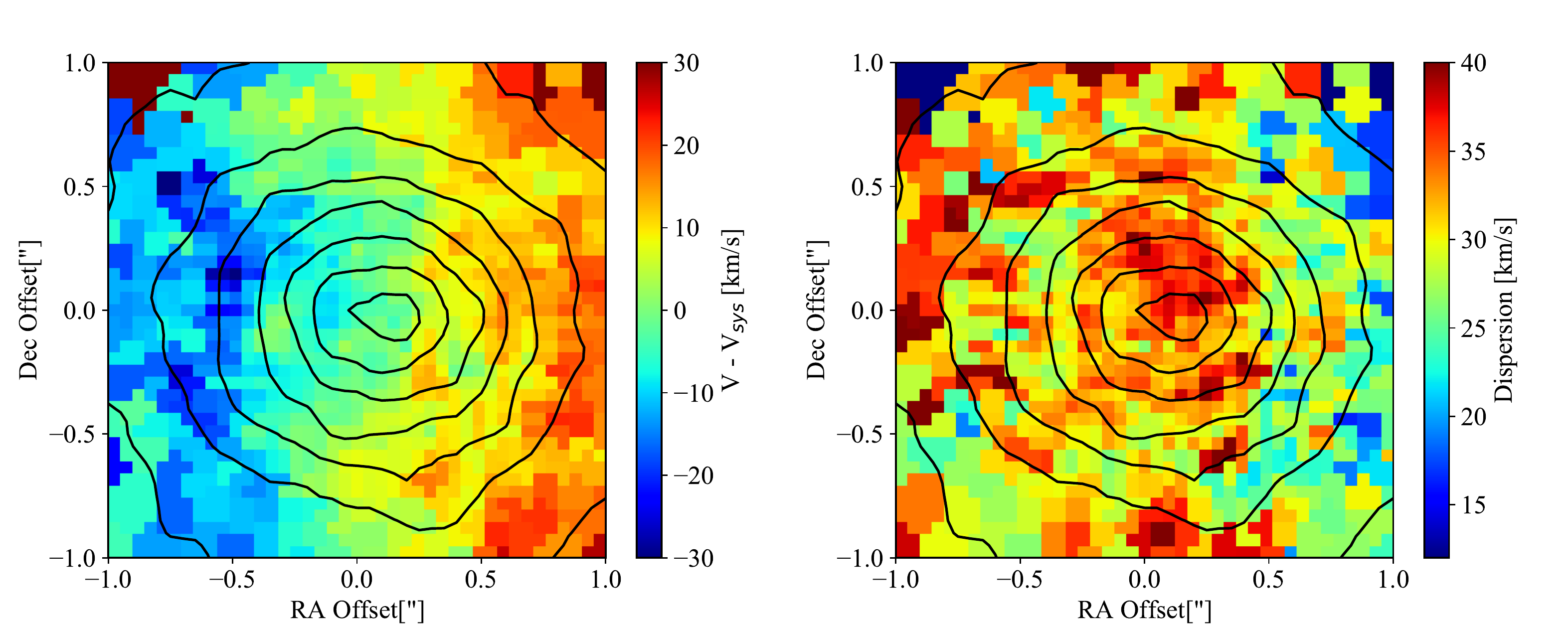}
\caption{Kinematics of B023-G78. The above two panels are the stellar kinematic maps (velocity and velocity dispersion respectively) of the cluster derived from adaptive optics assisted Gemini/NIFS data. The systemic velocity (V$_{sys}$) was estimated to be $-$435 km/s. }

\label{fig:kinematics}
\end{center}
\end{figure*}
There have been claimed dynamical detections of IMBHs in the Milky Way; e.g.~in $\omega$~Cen \citep{noyola10,baumgardt17} ($\sim$4--5$\times$10$^4$~M$_\odot$), M54 \citep{ibata09}($\sim$10$^4$~M$_\odot$), NGC6388 \citep{lutzgendorf15}, but none of these have been proven, and none are supported by evidence for accretion despite very deep radio searches that would be expected to detect even quiescent IMBHs \citep{tremou18}. In M31, one of the brightest clusters, G1, has been suggested to contain an IMBH of $\sim$2$\times$10$^4$~M$_\odot$ \citep{gebhardt02,gebhardt05}; however this detection is also controversial \citep{baumgardt03}, and a lack of accretion evidence was shown in \citet{miller_jones12}. 

Yet, despite the challenges of IMBH detection in GCs, it appears that at least some IMBHs do exist \citep[see recent review by][]{greene20}.  The most convincing detection of an IMBH is the bright, off-nuclear X-ray source HLX-1.  This object, found $\sim$3~kpc from the center of a massive galaxy has an estimated BH mass of a few $\times$~10$^4$~M$_\odot$ \citep{davis11,webb12,godet12,straub14}. This source appears to be surrounded by a star cluster as well \citep{farrell14}.  
   
In this paper, we use high-resolution mass models and kinematics to present the detection of a $\sim$10$^5$~M$_\odot$ IMBH with $>$~3$\sigma$ significance in B023-G78.  This cluster is the most massive GC in M31, with a dynamical mass of 6.8$^{+0.7}_{-0.6}\times$10$^6$~M$_\odot$ and a central dispersion of 33.0$\pm$1.8 km/s \citep{strader11}, and is located along the minor axis of M31 at a projected distance of 4.4~kpc towards its center (Figure~\ref{fig:kinematics}). Line index measurements by \citet{caldwell11} suggest a metallicity [Fe/H] = -0.7, while analysis of the width of the RGB suggests a significant metallicity spread \citep{fuentes_carrera08}. The reddening is uncertain due to a dust lane passing in front of this GC with values ranging from 0.23 -- 0.43, we use the E(B-V) value of 0.23 \citep{jablonka92} as our default value. We also assume the values A$_{F814W}$/A$_V$~=~0.59 and A$_{F606}$/A$_V$~=~0.91. Surface brightness profile fits performed by \citet{barmby07} using a single King profile suggests a core and tidal radius of 1.35~pc and 37.15~pc, respectively (and thus an effective radius of 3.7~pc/1.0"). We assume the distance of M31 (and also to B023-G078) to be 0.77~Mpc \citep{karachentsev04}. All the magnitudes in this paper are expressed in Vega magnitudes.

In \S~2 we present the imaging and spectroscopic data. \S~3 and \S~4 describe the surface photometry and the dynamical modeling performed on the cluster. \S~5 presents our discussion and conclusions.

\section{Data}
\subsection{HST Data}
We used archival $HST$ data for this cluster from the proposal ID:9719  (PI:~T.~Bridges)\footnote{The specific observations can be accessed via \href{https://mast.stsci.edu/portal/Mashup/Clients/Mast/Portal.html?searchQuery=\%7B\%22service\%22:\%22DOIOBS\%22,\%22inputText\%22:\%2210.17909/t9-pm76-g165\%22\%7D}{ doi:10.17909/t9-pm76-g165}}. The observations were performed with the Advanced Camera for Surveys/High-Resolution Camera (ACS/HRC) in the filters F814W and F606W. The exposure times were 2860~s and 2020~s respectively. 

The ACS/HRC has a pixel scale of 0.025$''$/pixel and a field of view of 29$''\times$26$''$. We downloaded the individual \texttt{.flt} files from Mikulski archive for space telescopes (MAST) and drizzled them using \texttt{Astrodrizzle} \citep{drizzlepac} to create the final image. The \texttt{MDRIZSKY} keyword was set to zero to avoid over-subtraction of the sky in the final drizzled image. The final color image (F606W - F814W) is shown in the right panel of Figure~\ref{fig:m31}.

The drizzled PSF in each band was obtained by inserting \texttt{Tiny Tim} PSFs into mock \texttt{.flt} images and drizzling them the same way as the science image. This procedure is similar to the one described in \citet{pechetti20}. We note that the F814W PSF has a significant amount of light in a large halo; a PSF of radius 5" was used to account for this.

\subsection{Gemini/NIFS Data and Kinematics}
 We obtained Gemini/NIFS laser guide star adaptive optics observations of B023-G78 on Oct.~7 and Nov.~9 2014 as part of program GN-2014B-DD-2 (PI:~A.C.~Seth). The data provides integral field spectroscopy in the $H$ band (1.48-1.79$\mu$m) over a field of view of 3$''$ (11 pc at 0.77 Mpc).  For our final data cube with a spaxel size of 0.05$''$, we combined the 6/8  900s dithered exposures using the Gemini IRAF packages, with modifications as described in  \citet{seth10,ahn18}. Despite the use of offset sky exposures, additional on-chip sky subtraction was required before combination, using the corners of the chip; this makes our useful field of view $\sim$2$''$.  The line spread function was measured from sky lines in each pixel with a median FWHM of 3.27\AA. The kinematic maps are shown in Figure~\ref{fig:kinematics}.
 
 Due to the data's high-spatial-resolution, we were able to resolve individual stars in the GC. To mitigate shot-noise effects due to the brightest cluster stars, we used the \texttt{PampelMuse} software \citep{kamann13} to generate a star subtracted cube of the GC. To describe the method in short, we fitted a single S\'ersic image to the continuum image of the cluster and inspected the fit residuals. We then manually identified the locations where the residuals suggested the presence of resolved stars and the resulting list was used to recover their PSFs as a function of wavelength using \texttt{PampelMuse} \citep{kamann13}. These PSF and the positions were used to extract the spectra of the input stellar sources. Finally, we combined the wavelength-dependent PSF model with the positions and the extracted spectra of the resolved stars to subtract their contributions from the NIFS data.
 

In deriving the stellar kinematic maps, we first performed Voronoi binning using the code from \citet{cappellari03}.
 After resampling the integral field spectroscopic data into bins of S/N~=~50, we estimate the kinematics in each bin using the penalized pixel fitting (pPXF) algorithm and code as described in \citet{cappellari17}. This code uses the full spectrum (1.5$\mu$ -- 1.8$\mu$) to fit the radial velocity (V), and velocity dispersion ($\sigma_e$).
  We used 65 Phoenix stellar templates from \citet{husser13}\footnote{http://phoenix.astro.physik.uni-goettingen.de/} with metallicities ranging from -1--0, log(g) from 1 -- 5.5, temperatures from 3600 -- 5500~K, and [$\alpha$/Fe] from 0 -- 0.4, covering the range of parameters expected to dominate the light in B023-G78. To estimate the kinematic errors, we performed Monte Carlo simulations by adding a random Gaussian error to the spectrum in each bin and re-fitting the kinematics. The standard deviation of those fits was taken as 1$\sigma$ uncertainties. The final derived kinematics are shown in Figure~\ref{fig:kinematics}. The central velocity dispersion is $\sim$37~km/s, while clear rotation is seen around the minor axis with an amplitude of $\sim$20~km/s and a systemic velocity of $\sim -$435~km/s. The integrated dispersion out to 1$''$ is 34.2~km/s; this is in reasonable agreement with the observed value of 31.7$\pm$1.7~km/s by \citet{strader11} using higher spectral resolution optical spectroscopy at seeing limited spatial resolution.  
\begin{figure*}[!t]
    \centering
    \includegraphics[width=0.465\linewidth]{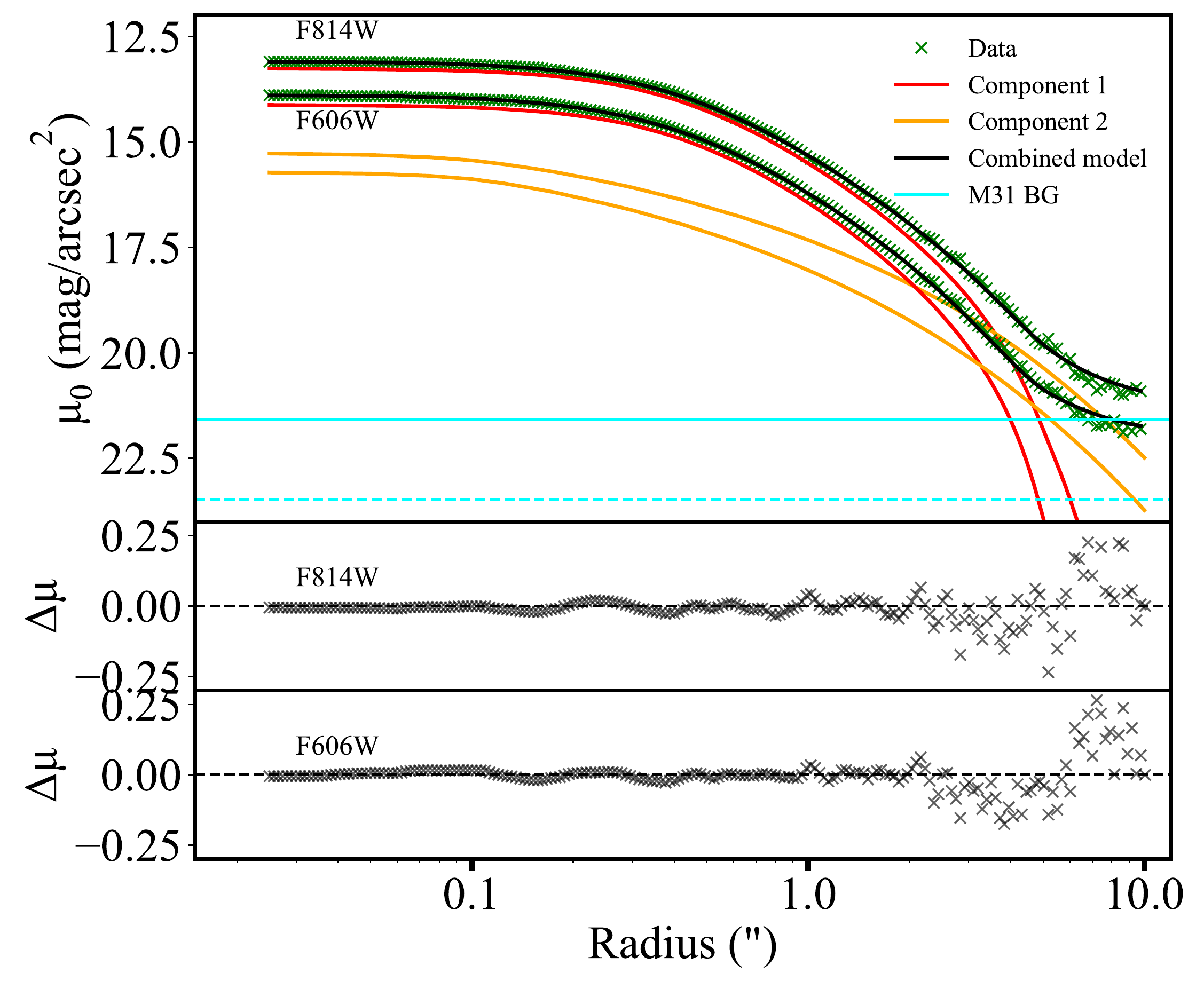}
     \includegraphics[width=0.475\linewidth]{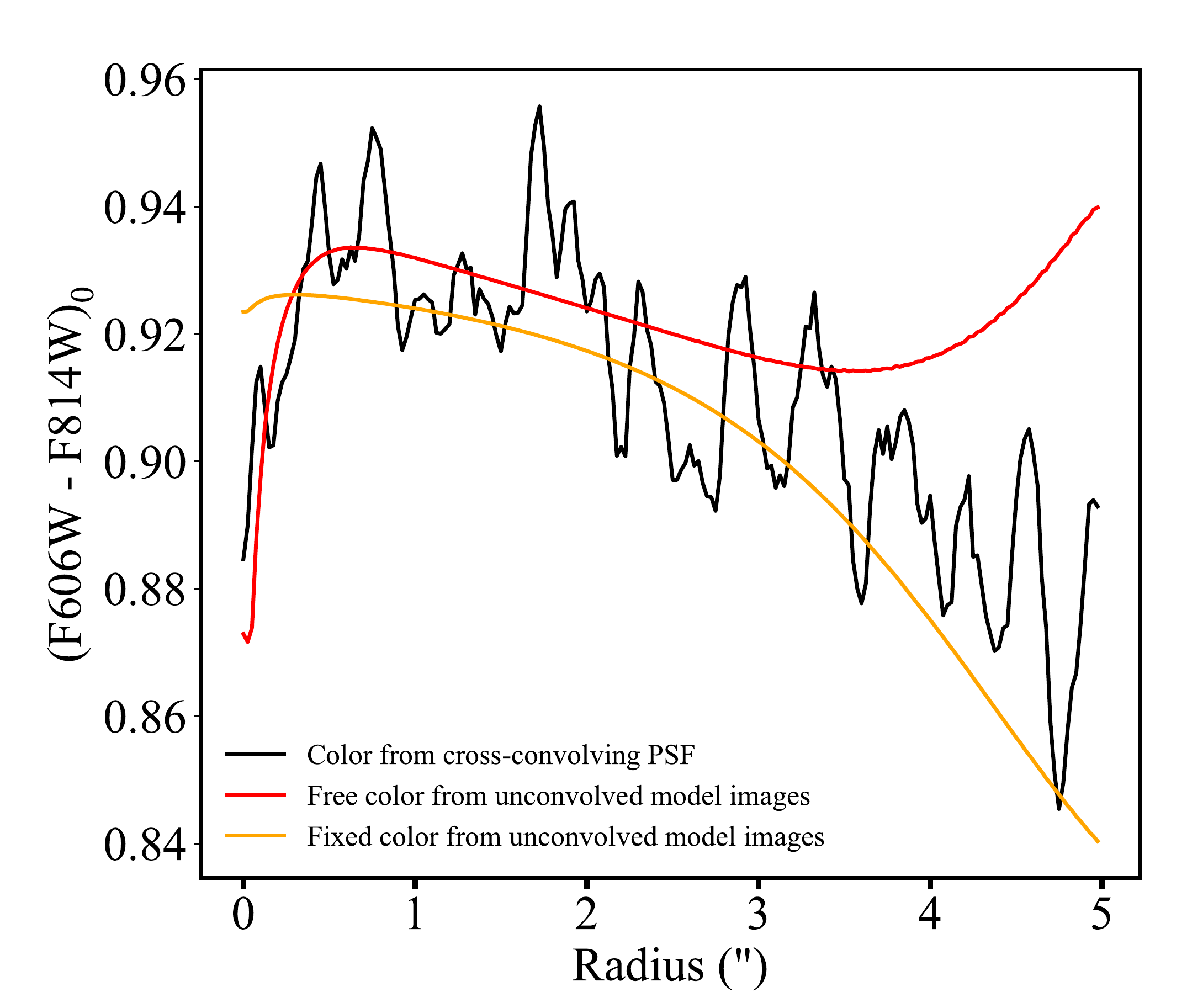}
    \caption{Surface brightness profile and color of B023-G78. {\em Left:} Best-fit model (King + S\'ersic) decomposition of the surface brightness of B023-G78 in F814W and F606W from the ACS/HRC data; note that surface brightnesses are in Vega magnitudes. The blue (solid and dashed) is the M31 background estimated using SDSS images (in F814W and F606W respectively; \S~\ref{sec:imfit}). The residuals (Data - Model) are shown in the bottom panels. {\em Right:} Color map derived by convolving the PSF of F814W to the F606W image and vice-versa (in black). The orange line is the color from the unconvolved model images, where the parameters of F814W were fixed to the best-fit model parameters of F606W, therefore requiring that the two components each have a unique color.  The red line is the color when the fit parameters in both the F814W and F606W were allowed to be free. The magnitudes in the right panel are extinction corrected with an E(B-V) of 0.23.}
    \label{fig:surf}
\end{figure*}
\subsection{Deriving the kinematic PSF}
\label{sec:kinematic_psf}
To perform dynamical modeling with precision, understanding the PSF of the Gemini/NIFS kinematic data is critical.
To determine the PSF, first, we astrometrically aligned a continuum image created from  the NIFS data cube (created without the additional on-chip sky subtraction) to the HST F814W image. Then, we used the HST F814W image within a 1$''$ radius and convolved it with a double Gaussian model of the PSF, varying the parameters of the Gaussians until the convolved HST image best matched the continuum image created from the data cube. We then convolved the resulting PSF model with the HST PSF to obtain the widths and relative strengths of the two Gaussian components. The best-fit FWHM of the inner and outer Gaussian component was found to be 0.127$''$ containing 31.4\% of the total light, and 0.58$''$ containing 68.6\% of the light, respectively. 

To account for systematic errors arising from the PSF model (described in \S~\ref{sec:systematic_errors}), we also created PSFs with different sets of inputs. We estimated a PSF as described above, but fitted the F814W image out to a larger radius (1.35$''$). Another PSF was generated as above but using the F606W HST image. The parameters of the PSFs in both the cases agreed within $\sim$10\% in the FWHMs and ratios of the components. This consistency is likely due to the lack of a strong color gradient seen in this cluster (see next section).  

\section{Creating Luminosity Models}
\label{sec:imfit}
 We fit the HST based surface brightness (SB) profiles of the cluster using the \texttt{IMFIT} code \citep{erwin15}. The code builds a 2-D model image using the input parameters and then convolves it with the given PSF. The best-fit $\chi^2$ is then estimated to find the closest model of the galaxy. Our best-fit model has two components, an inner King profile, and an outer S\'ersic profile.  The King profile \citep{king62} is described by the central intensity (I$_0$),  tidal radius (r$_t$), and core radius (r$_c$); although the \texttt{IMFIT} code uses a generalized King profile \citep{peng10}, we fix the power-law index $\alpha$ set to 2 to obtain the standard empirical King profile. The S\'ersic profile \citep{sersic68, graham05} is described by the S\'ersic index (n), effective radius (r$_e$), and the intensity at the effective radius (I$_e$). Apart from these input parameters to the models, we also provide position angle (PA), which is defined as the angle of the semi-major axis measured north through east counter-clockwise and ellipticity ($\epsilon$) as free parameters for each component to fit the 2-D image of the cluster. 

The image also includes background light from M31, which we assume is locally flat. We estimated the background from M31 and the sky levels of the HST images (which are not sky subtracted) by matching the SBs of the large-scale SDSS images in $r$ and $i$ bands transformed to HST F606W and F814W Vega magnitudes. After fitting for the HST image background levels, the transformed SDSS and HST SB profiles matched well beyond a radius of 5$''$.  To obtain the M31 background we used the mean surface brightness value in the region of 12--18$''$ from the SDSS image and incorporated this as a flat background component in our models. We estimated this background level to be at 21.57 and 22.59 mag/arcsec$^2$ in F606W and F814W respectively. The standard deviation in the SDSS surface brightness at large radii was taken as the 1$\sigma$ error on this estimate; $\sim$0.05 mags in both bands.

The SB profile fits were performed separately in both the F814W and F606W filters. Here, the fitting parameters were allowed to vary. To estimate the change in the color of the cluster in the model images, we fixed the best-fit model parameters of the F814W image to that of the F606W image; we found an (F606W-F814W)$_0$ of 0.94 mags for the inner component and 0.89 mags for the outer component. A 1-D radial profile of the cluster's surface brightness and the model fit is shown in the left panel of Figure~\ref{fig:surf}, which was derived from summing up the 2-D image in annuli of increasing radii. The best-fit parameters are given in Table~\ref{table:params}. We note that \citet{barmby07} show that a single Wilson profile provides a good fit to the 1-D profile of B023-G78.  However, the combination of the outer component's bluer color and its significantly higher ellipticity indicates a real physical difference in the two components.

As noted earlier, the F814W PSF has a red halo, and thus to examine the color profile in more detail, we created a cross-convolved color map, i.e.  we convolved the F814W image with the PSF of F606W and vice versa. The resulting color profile is shown in the right panel of Figure~\ref{fig:surf}. We show the central 5$''$, out to the radius of our PSF.  The observed color gradient roughly matches the expectations from our fixed parameter fits, with a $\sim$0.05 mag decline between the central arcsecond and 5$''$.
 Because this color gradient is so small, especially over the area we are fitting, we assume a constant M/L in our dynamical models, but we also explore mass models with a varying M/L in \S~\ref{sec:systematic_errors}. The theoretical color for a population of 10 Gyr and [Fe/H] of -0.7 is $\sim$0.82 mags using the PARSEC models \citep{bressan12}.  This is considerably bluer than the observed and model colors, suggesting that our assumed reddening, E(B-V)=0.23 \citep{jablonka92}, may be underestimated.  We discuss this further in \S~\ref{sec:extinction}.  

\begin{table}
\centering

\caption{Best-fit parameters in F814W and F606W for B023-G78}
\label{table:params}
\def\arraystretch{1.2}
\begin{threeparttable}
\begin{tabular}{ccccc}
\hline\hline
& Function & Parameter &  Best-Fit value  & Best-Fit values		\\
&        &              &   F814W     & F606W    				\\
\hline	
& King  & log~I$_0$     	&  	4.91 L$_\odot$/pc$^2$	&  	4.65 L$_\odot$/pc$^2$				\\
&		& r$_c$     	&  2.69~pc  				&   2.68~pc \\
&  		& c $=$ log(r$_t$/r$_c$) 	&  1.11					&  	1.12\\
& 		& $\epsilon$    &  0.10					&  	0.11	 \\
& 		& PA     		&  	80.0					&  	76.4\\
& 		& mag$_{tot}$     		&  	13.02					&  	14.22 \\
\\
& S\'ersic & log~I$_e$     	&  	1.79 L$_\odot$/pc$^2$	&  	1.24 L$_\odot$/pc$^2$				\\
&  		   & r$_e$     	&  18.74~pc   				&  	15.06~pc   							\\
&  		   & n    		&  2.56	  					&   2.52  								\\
&		   & $\epsilon$ &  	0.24					&  	0.26								 \\
&		   & PA     	&  	77.0					&  	85.6	\\
& 		& mag$_{tot}$     		&  	14.08					&  	15.24 \\
\\
& M31 & 	&  	21.57	& 22.59 \\
& background & & mag/arcsec$^2$ & mag/arcsec$^2$\\
\\
& Half-light & r$_{hl}$ & 4.23 pc & \\ 
&radius & & &\\
\hline
\end{tabular}
\begin{tablenotes}
$Note$: The cluster is fitted by a King + S\'ersic model. The parameters and their corresponding best-fit values are shown here. These parameters are from the free models. The magnitudes and luminosities are not extinction corrected.  
\end{tablenotes}
\end{threeparttable}
\end{table}

\begin{table}
\centering

\caption{Best-fit MGE parameters of F814W fits for B023-G78}
\label{table:mgeparams}
\def\arraystretch{1.2}
\setlength{\tabcolsep}{15pt}
\begin{threeparttable}
\begin{tabular}{ccc}
\hline\hline
 Intensity &  Gaussian width  & Axial ratio		\\
        (L$_\odot$/pc$^2$) &   (arcsec)     &     				\\
\hline	
24526    &  0.19  & 0.90 \\ 
2307  &    0.24 & 0.90 \\
43610  &     0.40 & 0.90 \\
11092  &     1.06 & 0.90 \\
4396  &   0.07 & 0.76 \\
3435   &   0.13  & 0.76 \\ 
2278  &    0.23 & 0.76 \\
694  &    0.29 & 0.76 \\
1478  &    0.39 & 0.76 \\
392   &     0.46 & 0.76 \\ 
1169  &    0.62 & 0.76 \\
805  &    0.90 & 0.76 \\
252  &     1.14 & 0.76 \\
190  &      1.34 & 0.76 \\
454  &     1.67 & 0.76 \\
288  &       2.80 & 0.76 \\
1.40  &     3.13 & 0.76 \\
84.4  &      5.77 & 0.76 \\
\hline
\end{tabular}
\begin{tablenotes}
$Note$: The MGE parameters for the fits to the F814W data in Table~\ref{table:params}. 
\end{tablenotes}
\end{threeparttable}
\end{table}


We use multi-Gaussian expansion (MGE) models to deproject the SB profiles for use in dynamical models. This method is described in detail in \citet{pechetti20}. In short, we used the best-fit parameters from Table~\ref{table:mgeparams} and converted them to MGE models using the \texttt{mge\_fit\_1d} code \citep{cappellari02}, sampling the SB profile logarithmically.  The final model contains 18 Gaussian components as described in Table~\ref{table:mgeparams}. The ellipticities from the \texttt{IMFIT} models were converted to axial ratios to deproject these MGEs. 
\begin{figure*}
    \centering
    \includegraphics[width=0.9\linewidth]{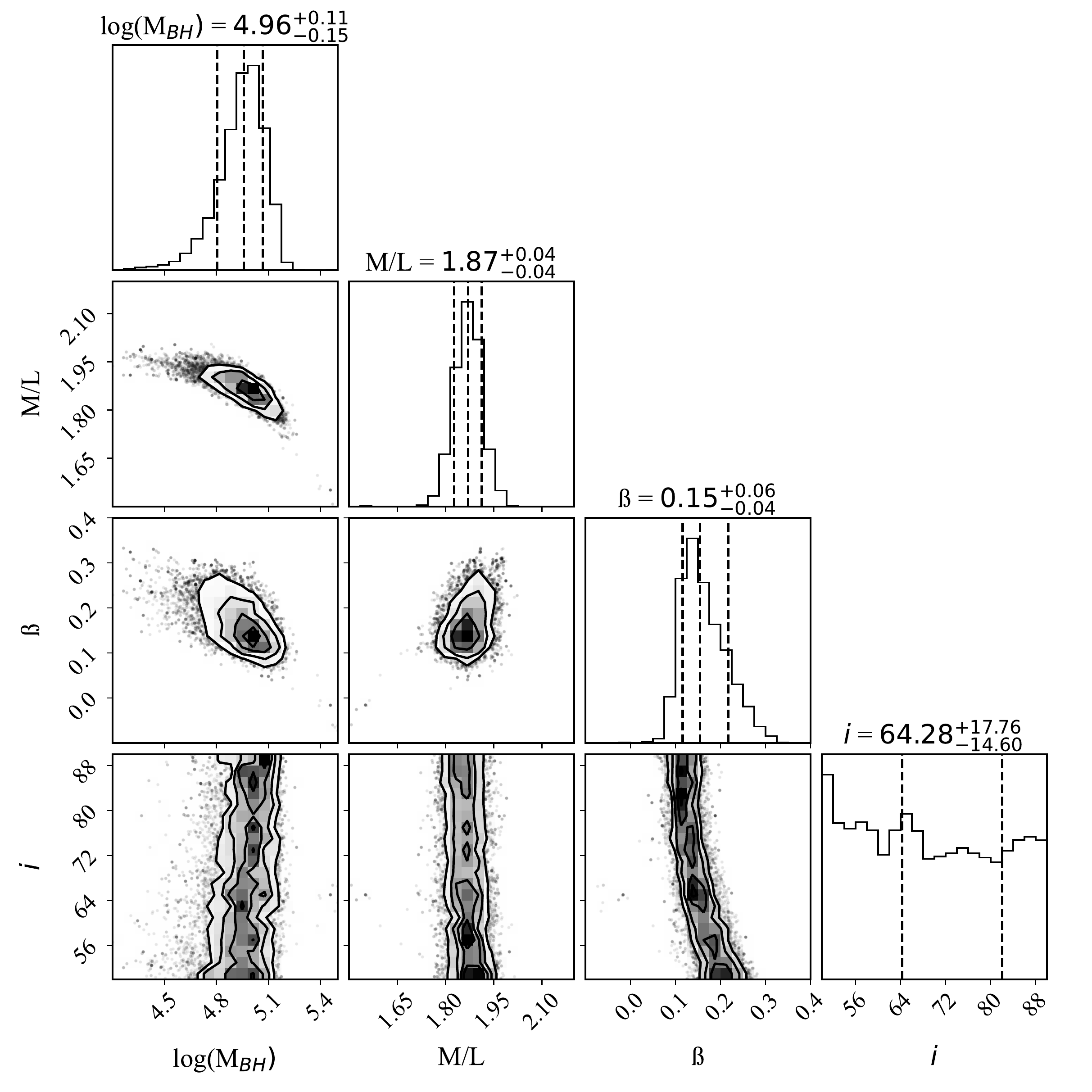}
    \caption{The output of JAM models from MCMC simulations showing the best-fit BH mass. $M_{BH}$ gives the black hole mass, M/L indicates the mass-to-light ratio in the F814W band, $\beta$ shows the anisotropy, and $i$ gives the inclination.  The top panel shows the probability distribution function of the black hole mass marginalized over all other parameters.}
    \label{fig:mcmc}
\end{figure*}
\section{Dynamical Modeling and BH Mass Estimates}

 In this section, we present dynamical models of B023-G078 that focus on constraining the mass of a possible central BH mass using
Jeans' anisotropic modeling \citep[JAM;][]{cappellari08}. 
We first present results for our default model, then explore the impacts of the uncertain extinction correction and possible systematic errors on our best-fit models. We present additional dynamical models exploring the possibility of a cluster of stellar mass black holes in \S~\ref{sec:stellarbh}.


\subsection{Results from Jeans Anisotropic modeling}

For estimating the BH mass, we used the JAM method for our dynamical models. These models use the 3-D deprojected MGE densities that were derived from the HST data in the previous section to create a gravitational potential. To this potential, a BH assuming a Gaussian potential with a very small scale ($\sim$0.01$''$) is added. Using the potential and MGEs, the Jeans' equations are solved to estimate an intrinsic value of the root mean square (RMS) velocity ($V_{RMS} = \sqrt{(V-V_{sys})^2+\sigma_0^2}$), where V is the rotation velocity, V$_{sys}$ is the systemic velocity, and $\sigma_0$ is the velocity dispersion. The estimated V$_{RMS}$ is then integrated along the line of sight to compare with the observed RMS velocities derived from the Gemini/NIFS data out to a radius of 1$''$.  Our default model uses the kinematic PSF derived from fitting the Gemini/NIFS data to the F814W image, the best-fit two-component King+S\'ersic model derived from the F814W image (Table~\ref{table:params}) , and the kinematics data cube after star subtraction. We discuss additional models used to assess our systematic errors in \S~\ref{sec:systematic_errors}.

We explore our JAM model fits by varying the following 4 free parameters: mass-to-light ratio (M/L), inclination angle ($i$), anisotropy parameter $\beta$, and BH mass $M_{BH}$, since they are degenerate.  We estimate the best-fit values by sampling the parameter space using Markov Chain Monte Carlo (MCMC) simulations with the \texttt{emcee} package \citep{foreman13}. We ran our models for 10000 iterations. 
The resulting posterior probability distribution functions of our model parameters are shown in Figure~\ref{fig:mcmc}.  

\begin{figure*}
    \centering
       \includegraphics[width=0.505\linewidth]{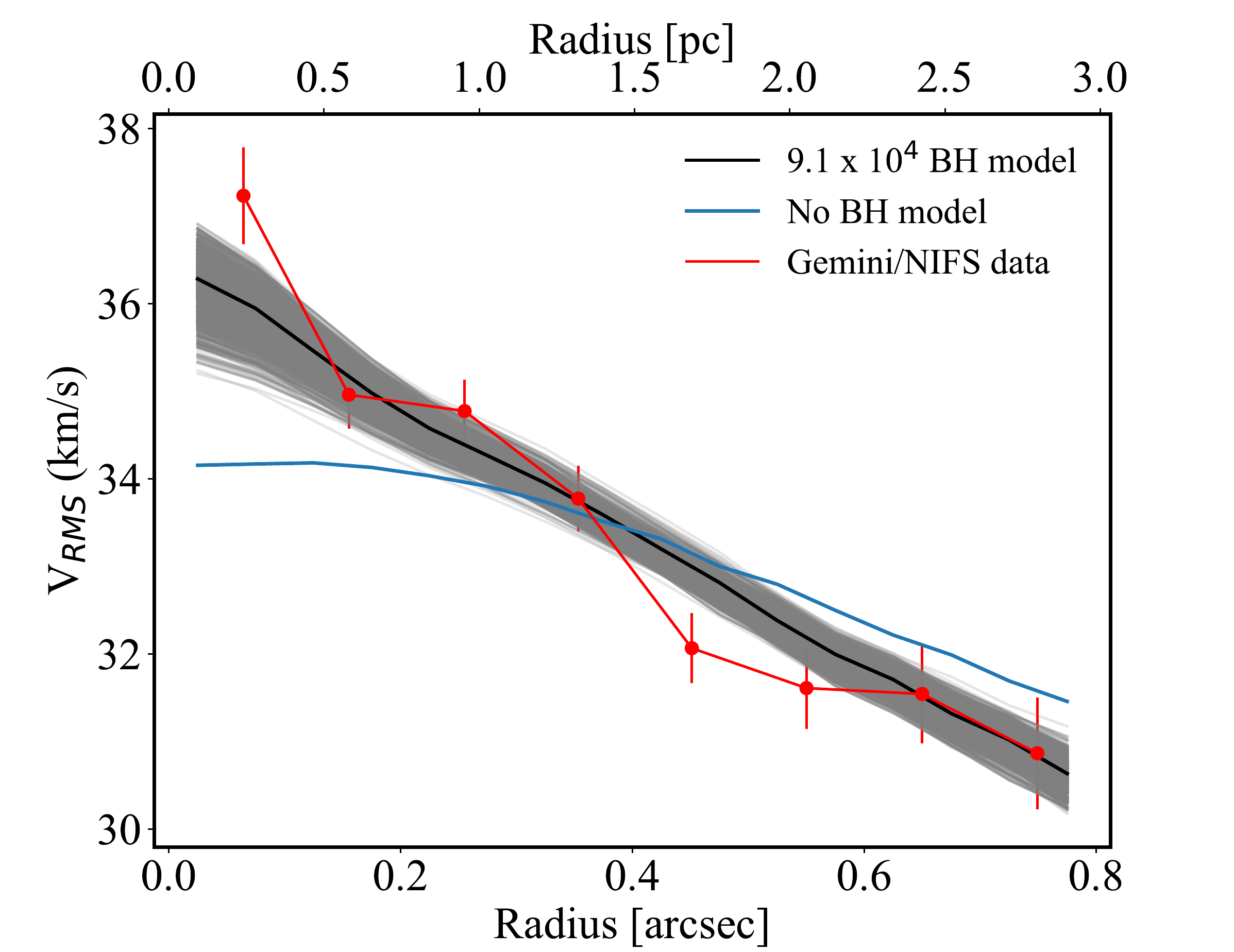}
     \includegraphics[width=0.49\linewidth]{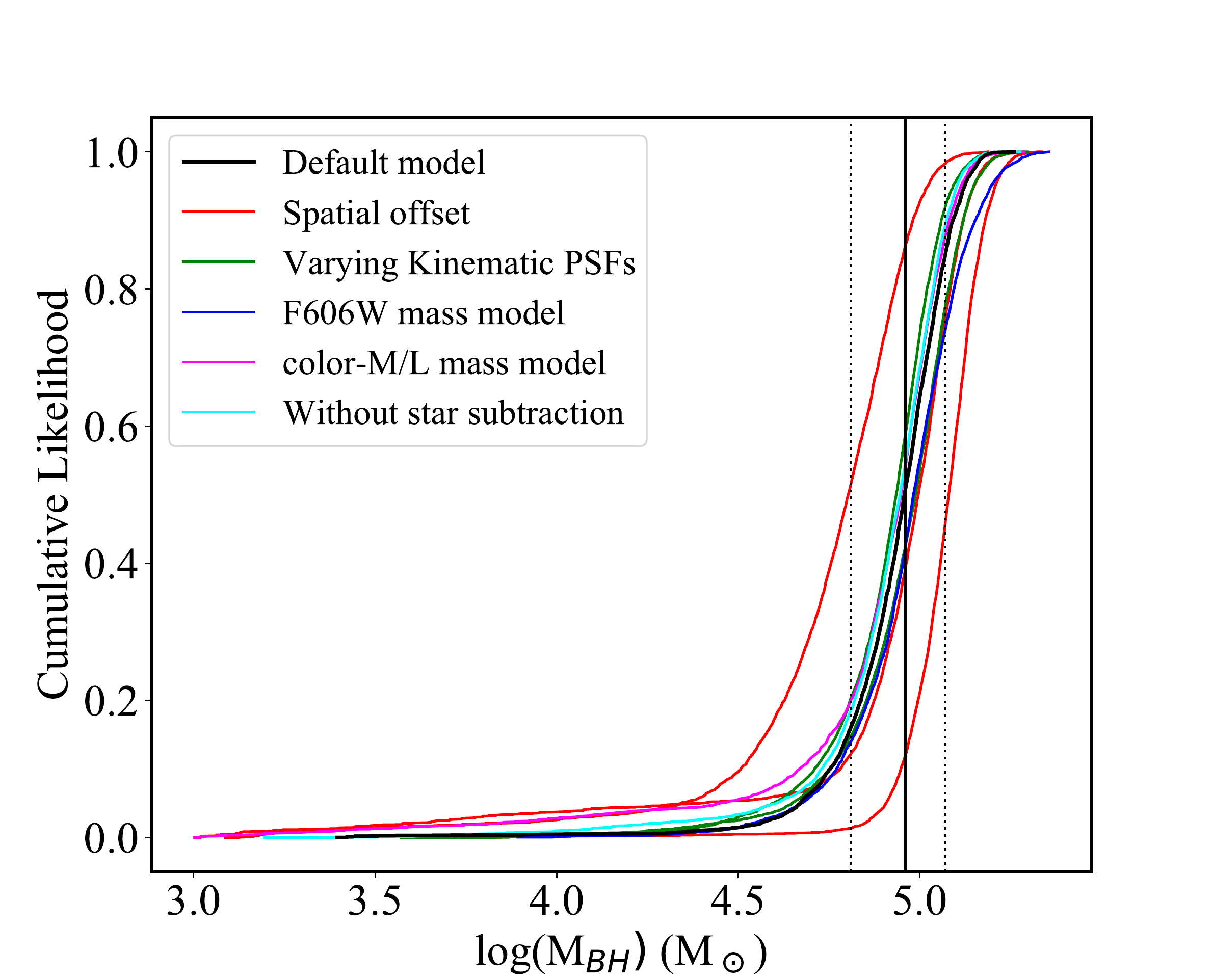}
    \caption{{\em Left:} A 1-D profile of the observed kinematics compared to model fits. Red points show annular V$_{RMS}$ values from Gemini/NIFS.  The black line shows the best-fit BH mass model with gray lines showing other models from the MCMC model fits.  The blue line is the best-fit no-BH model derived from fitting only 3 parameters (M/L, $i$ and $\beta$).  {\em Right:} Cumulative likelihood of the BH mass estimate. The default model is the black solid line. The dashed lines are the 1$\sigma$ uncertainty levels. We also show other models to highlight the level of systematic error in our default model.  This includes mass model variations (i.e. using the F606W image fits and varying the M/L based on the color of the components), spatial offsets of one NIFS pixel, fits to kinematics derived from data cubes without star subtraction, and fits where we vary the kinematic PSF. All but the spatial offsets result in changes at the $<$1$\sigma$ level of the black hole mass, while the spatial offsets (which are significantly larger than our uncertainties in the center) are consistent within 2$\sigma$ with the black hole mass in the default model.  }
    \label{fig:jam1d}
\end{figure*}

We obtain a best-fit BH mass of 9.1$^{+2.6}_{-2.8}\times$10$^4$~M$_\odot$. The $\chi^2$ of the best-fit model is 404.  The best-fit no-BH model has a $\Delta \chi^2$ of 30, excluding this model at $>3\sigma$ significance relative to the model with a BH.  We estimated the Bayesian information criterion (BIC) for the best-fit IMBH model and the no-BH model. The $\Delta$BIC was 24, which provides strong evidence against the no-BH model. A $\Delta$BIC$>$10 supports strong evidence for one model over another \citep{kass95}. For the best-fit BH mass and $\sigma_e$ as the integrated velocity dispersion at $\sim$0\farcs5, the sphere of influence radius (SOI$= G M_{BH}/\sigma_e^2$) is $\sim$0.33~pc or $\sim$0$\farcs$09; for comparison, the PSF core sigma (FWHM/2.35) is 0$\farcs$055, thus the SOI is resolved by our kinematic data as expected given the $>$3$\sigma$ significance of the BH mass detection.  
The best-fit M/L$_{F814W}$ is 1.87$^{+0.04}_{-0.04}$, giving a total dynamical mass of 6.22$\times$10$^6$~M$_\odot$ for this cluster.  This total dynamical mass is similar to that found in \citet{strader11} (6.8$^{+0.7}_{-0.6}\times$10$^6$~M$_\odot$). We discuss the uncertainties in the M/L due to extinction in the next subsection but note that the dynamical mass is robust to changes in extinction. The models also suggest moderate radial anisotropy with a best-fit $\beta$ of 0.15$^{+0.05}_{-0.03}$. The inclination is not well constrained, but does not affect the estimates of other parameters. To visualize the models and data better, Figure~\ref{fig:jam1d} shows a 1-D radial profile of the measured annular $V_{RMS}$ and the $V_{RMS}$ model prediction, as well as showing the best-fit model without a BH.
The model 1-D profiles are estimated by creating radial bins from the 2-D model and taking the median for each bin along the major axis of the cluster. Every iteration of the MCMC simulation within 1$\sigma$ is also plotted, which is the shaded region.
\subsection{Effects Of Extinction}
\label{sec:extinction}
As noted in the introduction, the dust lane passing in front of this cluster makes the extinction of this cluster poorly known.  This uncertainty translates directly into an uncertainty in the M/L of the cluster, however, the BH mass and total dynamical mass of the cluster are not sensitive to changes in the extinction because these are constrained by the combination of the kinematics and the shape of the mass model.  To check for any extinction variations within the cluster, we averaged the color of the cluster azimuthally, but there were minimal variations ($<$1\%). This suggests that the extinction is nearly constant in the region we are modeling.  Our default reddening of E(B-V)$=$0.23 (corresponding to $A_{F814W} = 0.427$) from \citet{jablonka92} provides an M/L of $\sim$1.9; the range of $I$ band mass-to-light ratios observed in other M31 clusters is $\sim$1--2 for high mass clusters with similar metallicity \citep[calculated from][]{peacock10,strader11}, thus our derived value is reasonable, although a bit on the higher side.  Using a higher reddening value in our dynamical modeling, like the E(B-V)$=$0.43 from \citet{caldwell11} gives an M/L$_{\rm F814W}=$1.35, also still within the range of observed M/Ls. We also analyzed the resolved photometry in our HST data.  Comparison of this data to Parsec isochrones \citep{bressan12} suggests the CMD position of the RGB and red clump stars are consistent with E(B-V)$=$0.23, and rule out significantly higher reddenings.  We therefore use the \citet{jablonka92} value as our default value.  We note that our choice of reddening/extinction does not impact the dynamical estimates of the best-fit BH mass or the total stellar mass of the cluster.

\subsection{Sources of Systematic Error}
\label{sec:systematic_errors}

Several systematic errors can affect our dynamical models. We discuss each of these and summarize their effect on our estimated BH mass in the right panel of Figure~\ref{fig:jam1d}. The default model as mentioned in the previous section is depicted in the black line.

One source of uncertainty in dynamically modeling GCs is defining the center \citep[e.g.][]{noyola10,anderson10}. When determining the surface photometry, \texttt{IMFIT} fits the center along with the cluster surface brightness. The formal error on the center was 0.156 mas, which underestimates the true uncertainty.  We also estimated the center by running the ELLIPSE task using IRAF. We did not fix the center and estimated the center using ellipses with semi-major axes of 0.2$''$--3$''$. We then determined the standard deviation of the measurements, which was $\sim$12.5~mas. Given that this is $\sim$1/4th the size of the kinematic pixels,  this uncertainty has minimal impact on our dynamical models.  Despite the small apparent uncertainty in our center, we tested the impact on the BH mass by shifting the central position of the cluster by 0.05$''$ (1 NIFS pixel), in the x and y direction. The resulting variations in the cumulative distribution function of the inferred BH mass, shown as the red lines in Figure~\ref{fig:jam1d}, were fairly large but still within the 2$\sigma$ uncertainty of our default model (black line).  Note that to get the center of kinematics to match that of the HST data, during our PSF analysis, we obtain the best-fit astrometry matching our NIFS data to the HST images.  

Another major source of potential systematic error is the kinematic PSF that we derive using a double Gaussian profile. As described in \S~\ref{sec:kinematic_psf}, we estimated the PSFs using different fitting radii on the F814W image, and using the F606W image. The impact of the PSF on our results is shown with green lines in Figure~\ref{fig:jam1d}. 

The luminosity/mass models we use also have two types of uncertainties: (1) uncertainties in the parameterization of the SB profiles, and (2) the possibility that the M/L varies with radius, invalidating the mass-traces-light assumption in our first model.  This could be due to varying stellar populations in the cluster (which we explore here) or due to mass segregation (discussed in the next subsection).  To explore the size of uncertainties in (1) we use the best-fit F606W King+S\'ersic model; this is shown as blue line in Figure~\ref{fig:jam1d}, which again did not create much variation in the BH mass.  To explore (2), we performed tests by varying the M/L in our mass model instead of assuming a constant M/L. We assigned a M/L for the King and the S\'ersic components based on their integrated colors using the theoretical color--M/L relations from \citet{roediger15}. These were then used as an input in our JAM models, and a single mass-scaling factor was used in place of the M/L \citep[as in][]{nguyen18, nguyen19}. This did not create much variation in the BH mass (shown as the pink line in Figure~\ref{fig:jam1d}).  

Finally, we used the original kinematics data cube rather than the one after star subtraction to estimate the BH mass but there was not much variation observed (shown as a cyan line in Figure~\ref{fig:jam1d}). 
Overall, these tests suggest that the dynamical signature of the IMBH in B023-G78 is robust.  

Based on all the systematic errors that we explore, we find that none of them substantially change the estimated BH mass.

\section{Discussion}
We first discuss the evidence that B023-G078 is a stripped nucleus, and the interpretation of our BH results in that context. We then consider a collection of stellar-mass BHs as an alternative to the IMBH interpretation, and finish by examining B023-G078 in a broader context.

\subsection{Additional Evidence that B023-G78 is a Stripped Nucleus}
The presence of an IMBH might be expected in B023-G78 if it is a stripped nuclear star cluster (NSC) of a once more massive galaxy \citep[e.g.][]{pfeffer13}.  B023-G78 is the most massive cluster in M31 and an outlier in the M31 globular cluster luminosity function \citep{barmby01,strader11}.  In the Milky Way, there is strong evidence that some of the most massive clusters are stripped NSCs. $\omega$~Cen consists of complex stellar populations that cover a broad metallicity distribution. In addition, it has recently been suggested as the former core of Sequoia or Gaia-Enceladus galaxy \citep[e.g.][]{majewski12,myeong19,simpson20,pfeffer21}. The NSC of the tidally disrupting Sagittarius dwarf galaxy, M54, provides evidence that this stripping process is ongoing. It also shows complicated star formation history and a spread in metallicity and ages of the stars \citep[e.g.][]{sarajedini95,siegel07,alfaro-cuello19,pfeffer21}. All of the globular cluster formation mechanisms are unable to explain these observations \citep[e.g][]{bastianlardo18}.  Assuming B023-G078 is a stripped NSC, we estimate a galaxy progenitor stellar mass of 5.3$\times$10$^9$M$_\odot$ using the galaxy-NSC mass relation from \citet{neumayer20}; this relation has significant scatter, but most known NSCs of B023's mass are hosted in galaxies above 10$^9$~M$_\odot$.   In this range of galaxy stellar masses the black hole occupation fraction is high \citep{nguyen19,greene20}. 
\begin{figure}
\label{fig:zgradient}

     \includegraphics[width=\linewidth]{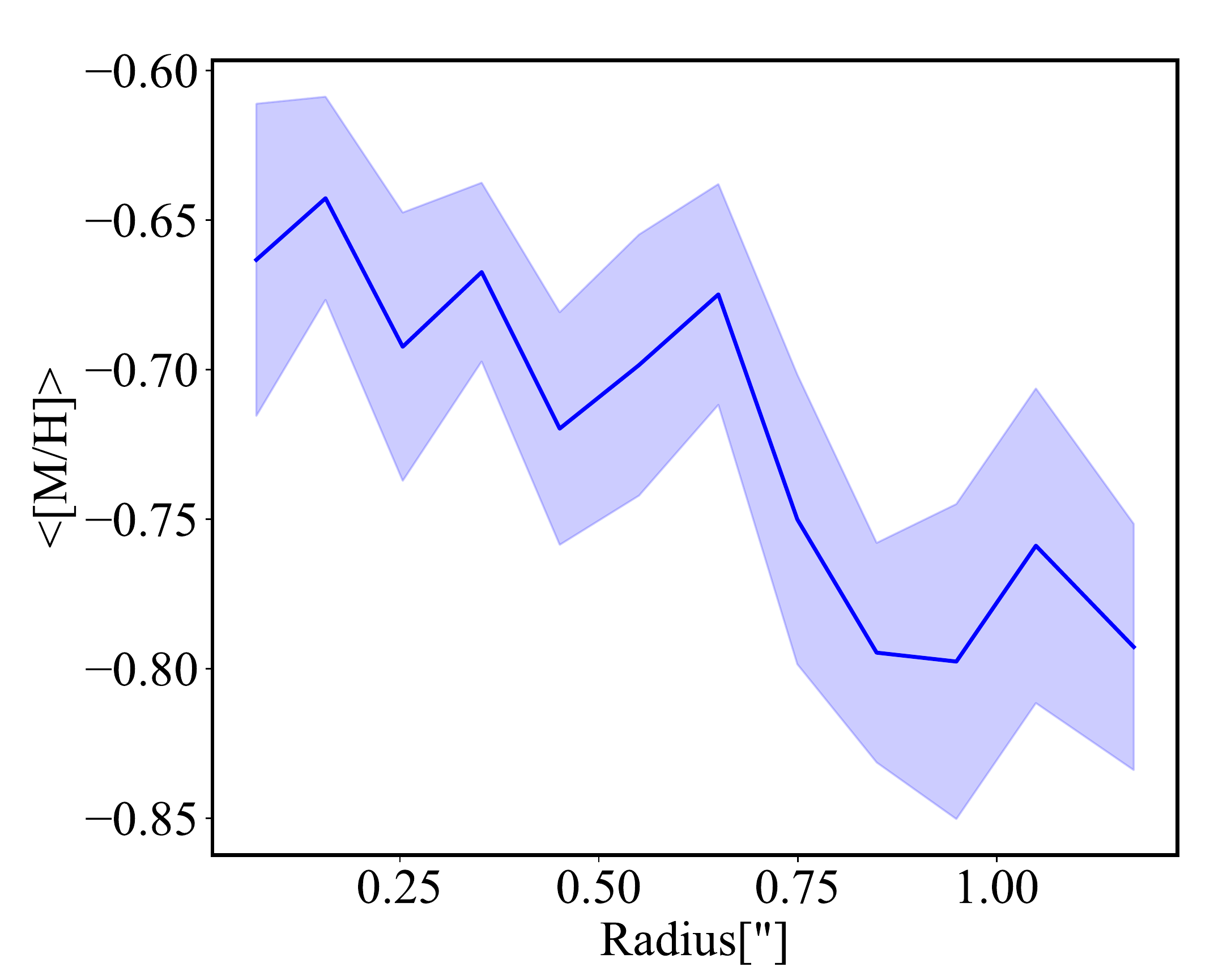}
    \caption{A clear metallicity gradient is seen in our Gemini/NIFS data.  We fit annular spectra using A-LIST models \citep{ashok21} to determine the light-weighted mean metallicity as a function of radius.  Error bars are based on Monte Carlo simulations.  
    }
\end{figure}

The metallicity \citep[e.g.][]{janz16} and metallicity spread \citep[e.g.][]{pfeffer21} of a globular cluster can provide additional evidence for a stripped NSC.  The metallicity of B023-G078 has been estimated to be roughly -0.7 from several studies of both spectra and CMDs \citep{fuentes_carrera08, perina09, caldwell11}.  
The observed metallicity for B023-G078 is within the observed range of NSC metallicities in its inferred host galaxy mass range \citep{neumayer20}.  Furthermore, the large spread in metallicities inferred from color-magnitude diagram modeling by \citet{fuentes_carrera08} provides strong evidence for B023-G078 being a stripped nucleus with a range of metallicities similar to M54 \citep{alfaro-cuello19} or $\omega$~Cen \citep{johnson10}.  The fits to APOGEE near-infrared spectra presented in \citet{ashok21} find a best-fit metallicity of -0.5$^{+0.3}_{-0.1}$, a best-fit [$\alpha$/M] of +0.1, and a considerably younger age ($\sim$6~Gyr) than any of the other 32 M31 GCs analyzed; this younger population also is suggestive of a stripped NSC where young populations are expected to form until the epoch of stripping \citep{neumayer20}.

Given our observed color gradient and the previously observed metallicity spread, we analyzed our NIFS spectra in radial annuli to detect any significant age or metallicity gradient in B023-G078. The spectra were binned into 12 annuli with a maximum radius of 1$\farcs$25 with $S/N$ ranging from $>$200 near the center to $\sim$80 at the largest radii.  We then fit the spectra using pPXF with the A-LIST spectral models, a set of simple stellar population templates created using APOGEE spectra \citep{ashok21}.  We selected Padova-based templates with [$\alpha$/M]$=$0.1, ages ranging from 2 to 12 Gyr, and [M/H] from -2 to +0.4.  The fits were very good, although due to the high $S/N$, the reduced $\chi^2$ was as high as 2.5 in the inner part of the cluster.  A light-weighted mean metallicity and age were calculated at each radius, and then a Monte Carlo analysis 
was run to determine the errors on these quantities (note that the error spectra were scaled by $\sqrt{\chi^2}$ of the best-fit at each radius during this analysis).  The light-weighted metallicity is consistent with previous metallicity determination and shows a clear negative gradient of $\sim$0.15 dex between the center and 1'' as shown in Figure~\ref{fig:zgradient}.  The light-weighted age is found to be 10.5$\pm$0.5~Gyr with no significant gradient.\footnote{This age is significantly older than the value measured by \citet{ashok21} using the same models but independent data.  We note that if we force a younger age on our fits, we get slightly worse fits, and higher metallicities consistent with the -0.5 found by \citet{ashok21}.} The lower metallicities at larger radii are also consistent with the bluer colors of our outer component inferred in our model fits to the B023-G078.  
The observed metallicity gradient is similar to those seen in $\omega$~Cen and M54 \citep[e.g.][]{suntzeff96,monaco05} with the metal-rich populations being more concentrated than the metal-poor populations.  Overall, we interpret the metallicity spread and gradient as evidence of the multiple generations of stars we expect to see in NSCs.

We note two additional pieces of evidence that favor B023-G078 being a stripped NSC.  First, the strong rotation (V/$\sigma$~$\approx$~0.8) seen is typical of NSCs \citep{neumayer20}, but is higher than those seen in Milky Way GCs \citep{kamann18}; note that this value is a lower limit due to the unknown inclination of the system.  Second, the two-component structure of the cluster is as expected from a stripped NSC \citep{pfeffer13}, and is similar to the more massive UCDs with known BHs \citep{seth14,ahn17,ahn18}.  The apparent (weak) color variation between the two components is also consistent with NSCs, where stellar population variations and gradients are expected \citep{neumayer20}.  Overall, there is strong evidence that B023-G078 is in fact a stripped nucleus from a galaxy in a mass range where central BHs are commonly found.  


\begin{figure*}
    \centering
  \includegraphics[width=\linewidth]{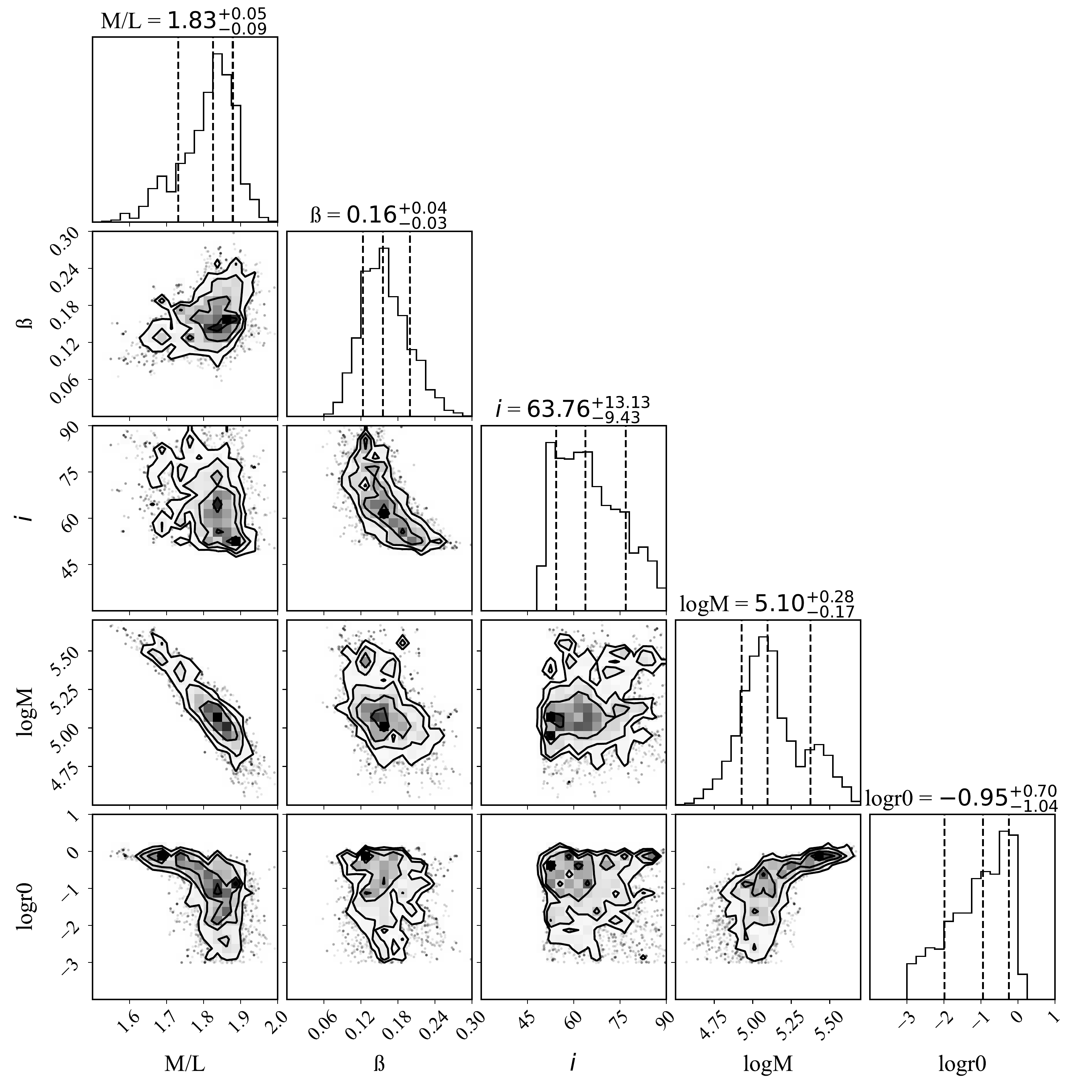}
    \caption{MCMC simulations of B023-G78 using a dark Plummer profile to describe a system of stellar mass BHs instead of an IMBH. The logM gives the total mass of the stellar mass BH subsystem, while logr0 indicates the Plummer radius, other parameters are the same as in Figure~\ref{fig:mcmc}.  The right two histograms give the best-fit "dark" component's mass and size marginalized over all other parameters.  The best-fit values of the total mass lie within 1$\sigma$ of the IMBH mass in Figure~\ref{fig:mcmc}, as do the inclination and anisotropy.}
    \label{fig:plummer}
\end{figure*}

\begin{figure*}
\label{fig:bhmass}
    \centering

     \includegraphics[trim={0cm 0cm 0cm 0cm},clip,width=0.7\linewidth]{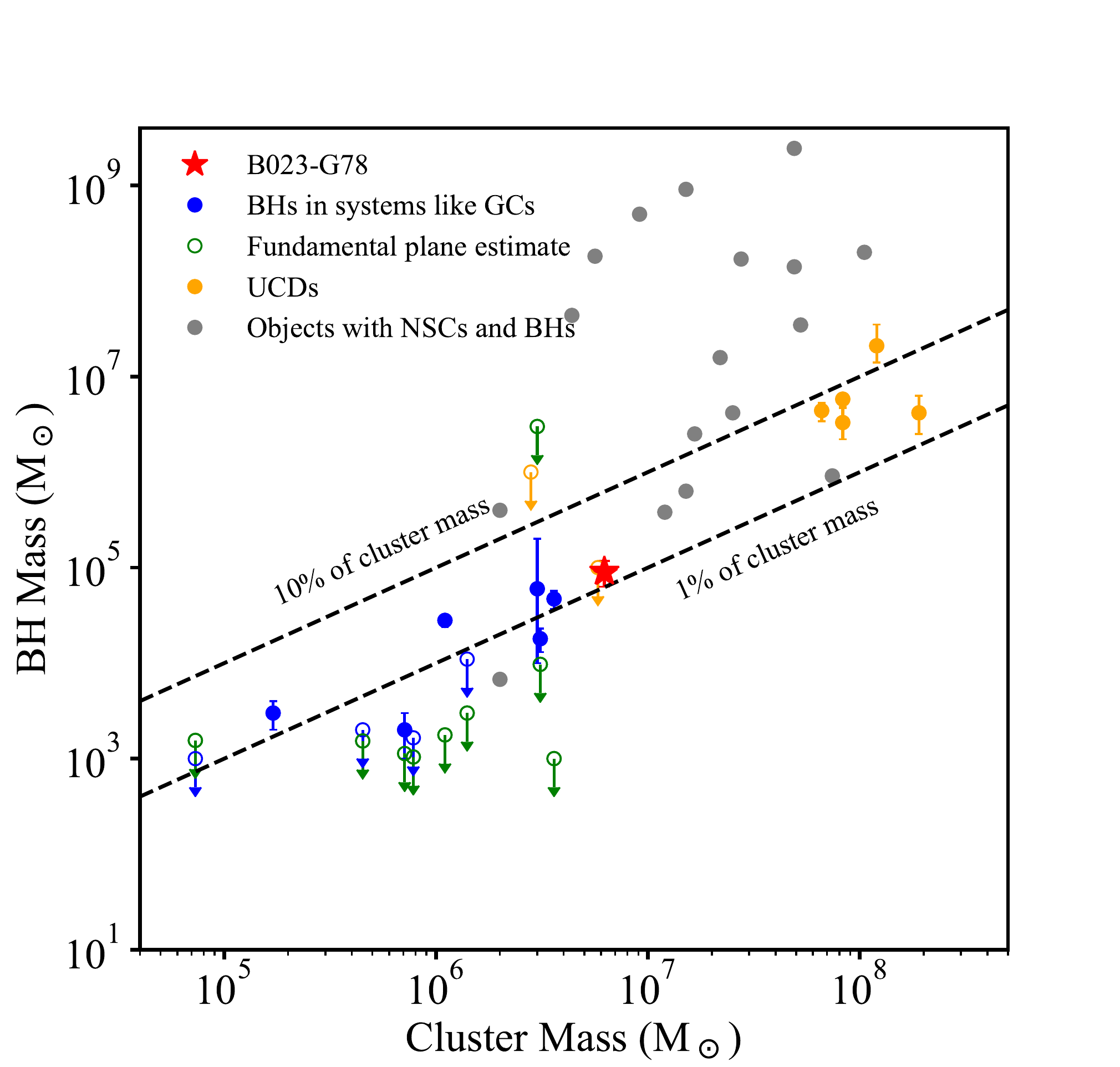}
    \caption{The black hole -- cluster mass diagram for systems comparable to B023-G078BH. The blue data points are BH mass estimates in GCs from the compilation of \citet{greene20}. The green data points are the fundamental plane upper limits of the same clusters from the same compilation. The orange points are stripped nuclei (UCDs) from \citep{seth14,ahn17,ahn18,voggel18,afanasiev18}. The gray points are the objects that have an estimate of both the BH and NSC mass from the compilation in \citet{neumayer20}. All the open circles are upper limits.  The dashed lines show BH masses that are 1\% and 10\% of their cluster mass. B023-G78 is the highest mass BH detected in a cluster below 10$^7$~M$_\odot$.}
\end{figure*}
\subsection{Possible alternatives to a central IMBH}
\label{sec:stellarbh}

 Dynamical evolution is expected to increase the M/L of clusters  both at the center, due to the mass segregation of BHs and neutron stars, and in the outer parts, due to kicks received by low-mass dwarf stars \citep[e.g.][]{denbrok14m15,baumgardt17}.  The mass segregation of the remnants happens on a timescale less than the half-mass relaxation time, which in B023-G78 is $\sim$14~Gyr, and thus it is expected that the BH subsystem will be mass segregated.  The expected mass fraction of stellar-mass BHs retained over time remains extremely uncertain due to poorly understood BH natal kicks from supernovae.  Observationally, constraints on the BH kicks derived from the 3-D velocities of X-ray binaries suggest typical kicks~$>$100~km/s \citep{atri19}, with a small fraction having much lower kick velocities; these are perhaps BHs formed from direct collapse.  The observed kicks are higher than expected from theoretical prescriptions that base the natal kicks on the better constrained neutron star kick distribution with a linear decrease in mass due to mass fall back and momentum conservation  \citep[e.g.][]{belczynski02,morscher15, banerjee20, mapelli21}.  The observed kick velocities are also above the escape velocities of even the most massive Milky Way clusters including $\omega$~Cen \citep{gnedin02}.  In addition to uncertainties due to kicks, additional uncertainty on the retention fraction of stellar mass-BHs comes from the unknown initial conditions for clusters including the high-mass stellar initial mass function \citep[e.g.][]{baumgardt18}, and the uncertainty in the initial-final mass relation of BHs \citep[e.g.][]{spera15,mapelli21}. 
 
 Models with $\sim$5\% of the cluster mass in segregated stellar mass BHs  are able to explain the rise in the central dispersion in $\omega$~Cen \citep{zocchi19} and may be preferred to an IMBH model due to the lack of a high-velocity tail in the individual stellar velocities near the center \citep{baumgardt19}. An alternative constraint on BH mass fractions in Milky Way clusters was made by \citet{weatherford20} through modeling the observed mass segregation of stars. They constrained the BH mass fraction in Milky Way GCs, and found them to be $<$1\% in 48/50 clusters (including the massive clusters 47 Tuc and M54). They report a correlation between the BH mass fraction and the ratio of the core radius to the half-light radius, and find two clusters with $r_c/r_{hl} > 0.75$ to have BH mass fractions of up to 2\%. B023-G078's large $r_c/r_{hl}$ thus suggests a high mass fraction of BHs may be present.  We note that the tidal stripping of star clusters can also lead to very high BH mass fractions as stars are lost from the cluster faster than the mass-segregated BHs \citep{gieles21}.

Relative to $\omega$~Cen, the higher metallicity of B023-G078 should lead to higher BH natal kicks (potentially lowering retention fractions) and lower typical BH masses and total BH mass fractions.  To get a sense of the potential maximum mass fraction in BHs, we assumed a \citet{kroupa01} IMF, the stellar evolution codes SSE \& BSE using an [Fe/H]$=-0.7$ \citep{hurley2000,hurley02} and the initial-final mass relations and BH kick prescriptions from \citep{banerjee20}.  This combination yields a total initial mass in BHs of 4.3\%, making up 7.8\% of the final mass. Removing BHs that receive kicks (and keeping only those that directly collapse), the present-day total mass fraction in BHs is 5.5\%. As noted above, the retention of BHs is highly uncertain, and the kick prescription used here doesn't match that of observed X-ray binaries \citep{atri19}.  On the other hand, the high mass of B023-G78 makes it plausible that a significant fraction of stellar-mass BHs  are retained \citep[e.g.][]{kremer20}.
Thus it appears possible that the inferred central IMBH in B023-G78 may instead be a collection of stellar mass BHs. We examine this possibility further below.

\subsubsection{Testing the Stellar Mass BH Scenario with JAM Models}
A collection of stellar-mass BHs differs from an IMBH because its mass distribution is extended, and this extent may be resolvable by our observations.  In $\omega$~Cen, the best-fit distribution of stellar-mass BHs from \citet{zocchi19} can be described as a Plummer density profile ($\rho = 3M(1+r^2/r_0^2)^{-5/2}/4\pi r_0^3$) with the ratio of the BH subsystem Plummer radius ($r_0$) to the cluster half-light radius ($r_{hl}),$ $(r_0/r_{hl}) \sim 0.3$.  This ratio would correspond to a $r_0$ of $\sim$1.3~pc (0$\farcs$3) in B023-G078; this is significantly broader than the core of our PSF and thus may result in measurable changes in our data relative to the point mass assumed in our IMBH models.  However, we note that in the distribution function-based models of \citet{zocchi19}, the amount of mass segregation between BHs and stars is fixed by a single parameter that is not well constrained, thus the ratio of $(r_0/r_{hl})$ is uncertain. A previous paper by \citet{breen2013} use theory, gas models, and $N$-body models on idealized clusters to understand the expected distribution of their BHs; they find that for the parameters of $\omega$Cen the ratio of half-mass radius of the BH sub-system ($r_{\rm h, BH}$) over $r_{\rm hl}$ is $r_{\rm h, BH}/r_{\rm hl} \simeq 0.15$. This ratio depends on the BH mass fraction; for a mass ratio of  $\sim$1\% as we find for the IMBH in B023-G078 they find $r_{\rm h, BH}/r_{\rm h} \simeq 0.1$.  For a Plummer profile this translates to $r_0/r_{hl} \sim 0.08$, which in B023-G078 would give an $r_0$ of just 0.3~pc, or 0$\farcs$09, only slightly larger than the PSF core Gaussian width of 0$\farcs$055 making for a more challenging measurement. Thus, if a significant BH subsystem is present in B023-G078, it is unclear whether we expect it to be significantly resolved by our observations.

To test whether an extended distribution of stellar mass BHs fits our kinematic data, we ran a new set of  JAM models replacing the central BH with a "dark" Plummer density profile.  
To include this in our JAM models, we created an MGE for the Plummer profile and included the Plummer radius ($r_0$) and the total mass (M) as free parameters in our MCMC simulations along with $M/L$, inclination, and $\beta$. The results are shown in Figure~\ref{fig:plummer}. From our simulations, we find that the median total mass of the dark component ($\sim$1.3$\times$10$^5$) is within $\sim$1$\sigma$ of our estimate for the IMBH. The median of the posterior for the Plummer $r_0$ parameter was $\sim$0.11~pc (0$\farcs$03) making it unresolved at our resolution; the best-fit value of 0.09~pc is also consistent with this small and unresolved $r_0$.  
The 95\% confidence upper limit on $r_0$ is 0.56~pc, thus the upper limit on $r_0$/$r_{hl}$ is 0.13. The mass of the dark system increases with increasing size, and for the $r_0$ upper limit, the corresponding upper limit on the total mass of the BH subsystem is 2$\times$10$^5$~M$_\odot$, 3.2\% of the total system mass.



We also estimated the BIC for the IMBH simulations and the models with the Plummer profile. We find a $\Delta$BIC of 6.3 providing positive evidence in favor of the IMBH models.  Combining this with the considerable evidence that B023-G078 is a stripped NSC,  we, therefore, favor an IMBH interpretation for our observations.  However, a compact system of stellar mass BHs is also a possible explanation for the observed rise in the central dispersion, as long as the $r_0 < 0.56$~pc and the total mass in the BH subsystem is $\lesssim$3\%.  

We note that it is possible that both an IMBH and a significant population of mass segregated stellar mass BHs are present.  A central BH significantly slows the process of mass segregation but does not completely halt it \citep{antonini14}.  While we do not model this hybrid case here, the constraints on the total mass of the dark Plummer model above likely give an upper limit on the mass of the stellar mass BH subsystem, even in the case of co-existence with an IMBH. We summarize the results and the properties of B023-G078 in Table~\ref{table:summary} and Table~\ref{table:b023}.

\begin{table}
\centering

\caption{Summary of results}
\label{table:summary}
\def\arraystretch{1.5}
\setlength{\tabcolsep}{6pt}
\begin{threeparttable}
\begin{tabular}{cccc}
\hline\hline
 &IMBH & No-BH  & Plummer		\\
& model & model & model \\
\hline	
M$_{BH}$ (M$_\odot$)   &  9.1$^{+2.6}_{-2.8}\times$10$^4$  & -- & 1.3$^{+1.1}_{-0.4}\times$10$^5$ \\ 
$M/L$ (M$_\odot$/L$_\odot$)  &    1.87$^{+0.04}_{-0.04}$ & 1.99$^{+0.02}_{-0.02}$ &1.83$^{+0.05}_{-0.09}$\\
$\beta$  & 0.15$^{+0.06}_{-0.04}$  &  0.22$^{+0.04}_{-0.04}$ & 0.16$^{+0.04}_{-0.03}$ \\
$i$  &  64$^{+18}_{-15}$  & 59$^{+16}_{-7}$ & 64$^{+13}_{-9}$ \\
$r0$ (pc) & -- &   --& 0.09\\ 
$r0$ upper limit (pc)& -- & -- & 0.56 \\
Best-fit $\chi^2$ & 404 & 434  & 404 \\ 
\hline
\end{tabular}
\begin{tablenotes}
$Note$: The best-fit parameters from the three different models we fit to the cluster.
\end{tablenotes}
\end{threeparttable}
\end{table}

\begin{table}
\centering

\caption{B023-G078 cluster properties}
\label{table:b023}
\def\arraystretch{1.2}
\setlength{\tabcolsep}{10pt}
\begin{threeparttable}
\begin{tabular}{lc}
\hline \hline
\\
Central V$_{RMS}$  &    37.2$\pm$0.6 km/s \\
V/$\sigma$  &   0.8 \\
Cluster mass &	6.22$^{+0.03}_{-0.02}\times$10$^6$M$_\odot$\\
BH Mass & 9.1$^{+2.6}_{-2.8}\times$10$^4$~M$_\odot$ \\
BH mass fraction & ~1.5\%\\
Half-mass relaxation time & ~14 Gyr\\
$[Fe/H]$  &   -0.65 (center) to -0.80 (at 1'')  \\
Age  & 10.5$\pm$0.5 Gyr \\
Assumed E(B-V) & 0.23    \\ 

\\
\hline
\end{tabular}
\begin{tablenotes}
$Note$: B023-G078 properties that we derived from our analyses. The E(B-V) value is from \citet{jablonka92} and used as a default value in this paper. 
\end{tablenotes}
\end{threeparttable}
\end{table}

\subsection{B023-G78 in Context}
Assuming our observed dynamical signature is an IMBH, we consider how it compares to other IMBH candidates and UCD/BH systems in Figure~\ref{fig:bhmass}.  At the lower mass end, a comparison sample of claimed dynamical detections of massive BHs in GCs, as well as published upper limits for the same clusters are shown from the recent compilation of \citet{greene20}.  We note many of the dynamical detections plotted here are disputed and refer readers to \citet{greene20} for details.  In addition we add higher mass UCDs from recent discoveries, as well as present-day galaxies with both NSCs and BHs to provide context. 

Relative to any other Local Group star cluster, the $\sim$9$\times$10$^4$~M$_\odot$ BH in B023-G78 is the highest mass detection claimed, double the suggested mass of the BH in $\omega$~Cen \citep[e.g.][]{noyola10}; as noted previously this IMBH detection has been contested \citep[e.g.][]{zocchi17,zocchi19,baumgardt19}.  It is also more significant than the $<$3$\sigma$ detection of a 2$\times$10$^4$~M$_\odot$ BH in G1 \citep{gebhardt05} derived from data with similar physical resolution.  

In comparison with the BHs previously found in other higher-mass UCDs, B023-G78 represents the first case in the IMBH regime, with all other BHs having both higher masses and mass fractions.  Relative to central BHs in present-day galaxies, the mass is the lowest dynamical estimate apart from the $\sim$10$^4$~M$_\odot$ BH suggested in NGC~205 \citep{nguyen19}.  The most comparable present-day NSC+BH system is NGC~4395, which hosts a $\sim$4$\times$10$^5$~M$_\odot$ BH, inferred both dynamically \citep{denbrok15} and from reverberation mapping \citep[e.g.][]{peterson05}, that lies in a $\sim$2$\times$10$^6$~M$_\odot$ NSC \citep{denbrok15}. The inferred IMBH in B023-G78 is also comparable to the lowest mass BHs inferred from accretion \citep[e.g.][]{baldassare15,chilingarian18}. 

We also checked for possible BH accretion signatures in B023-G078. There is no cataloged X-ray source matching the location of B023-G078 in the deep XMM mosaic of \citet{stiele11}. The faintest cataloged sources close to the location of B023-G078 have 0.5--4.5 keV XMM/EPIC unabsorbed fluxes of $2.1 \times 10^{-14}$ erg s$^{-1}$ cm$^{-2}$, which corresponds to a 0.5--10 keV X-ray luminosity of $1.9 \times 10^{35}$ erg $^{-1}$ assuming a photon index of $\Gamma = 1.7$. Hence the non-detection of B023-G078 in these data suggests a 0.5--10 keV upper limit of  $L_X \lesssim 2 \times 10^{35}$ erg $^{-1}$.  Using this upper limit to the X-ray luminosity in B023-G078 combined with our derived dynamical BH mass, the predicted 5 GHz luminosity is $<$~8.5 $\mu$Jy \citep{plotkin12}. B023-G078 is not detected in VLASS, and with an RMS noise in the VLASS image of 127 $\mu$Jy/bm, we can estimate a 3-$\sigma$ upper limit of $<$~381~$\mu$Jy. Therefore, in this case the X-ray limit (if accurate) is much more constraining than the radio data, although it would be possible to get significantly deeper radio data.  We also note that the presence of stellar mass black holes could also lead to detectable X-ray binaries, as B023-G078 does have a very high collision rate.  However, among the highest collision rate GCs in M31 only a fraction ($<$~half) appear to have bright X-ray sources \citep[e.g.,][]{peacock10}. We note in this context that in $\omega$~Cen, which as discussed above, may host a large cluster of stellar mass BHs \citep{zocchi19,baumgardt19}. However, no bright X-ray binaries are found, with the brightest X-ray sources being $<$10$^{33}$~ergs/s \citep{henleywills18}.

One potentially comparable systems detected via accretion is HLX-1, a bright off-nuclear X-ray source with an inferred BH mass of 10$^{4-5}$~M$_\odot$ \citep[e.g.][]{webb12}. Due to the light from HLX-1 itself, constraining the age and mass of the surrounding stellar cluster is challenging \citep[e.g.][]{soria10,farrell14,soria17}, but if it is old, its mass is estimated to be $\sim$3$\times$10$^6$~M$_\odot$ \citep{soria17}. 

\section{Conclusions}
We have presented adaptive-optics GEMINI/NIFS IFU kinematic data of M31's most massive star cluster, B023-G78.  We combined these data with mass models derived from HST ACS/HRC to constrain the mass content, including a possible central black hole in this massive star cluster. We find the following:
    \begin{enumerate}

    \item The kinematics of B023-G78 show a rise in the integrated velocity dispersion to $\sim$37 km/s, and a peak rotation of $\sim$20 km/s.  
    \item The surface brightness profile requires at least two components to fit, and shows a small color gradient, with the outer component being $\sim$0.05 mags bluer than the inner component.  A significant metallicity gradient of $\sim$0.15 dex is also seen within the central arcsecond.
    \item Our best-fit JAM dynamical models give a BH mass of 9.1$^{+2.6}_{-2.8}\times$10$^4$~M$_\odot$, M/L$_{F814W}$ of 1.87$^{+0.04}_{-0.04}$ and anisotropy 0.15$^{+0.06}_{-0.04}$.  The BH detection is highly significant $>$3$\sigma$, and systematic errors are $<$10\% on the best-fit BH mass.  
    \end{enumerate}

We discuss the possibility that this BH can be explained due to a collection of dark stellar remnants, and constrain the extent of these remnants and find the derived extent of the dark remnants are mostly unresolved by our observations, with an upper limit on the Plummer $r_0$ of 0.56~pc.  We favor the presence of a single IMBH given the other indications that B023-G78 is a stripped nucleus, as well as the apparent compactness of the dark component. Higher spatial-resolution data would give improved constraints on the nature of the central  dark mass and should be a high priority in the forthcoming era of extremely large telescopes.\\

The authors thank Mark Gieles and Alice Zocchi for useful conversations about this work.  R.P. and A.C.S acknowledge support from grants NSF AST-1350389 and AST-1813708. N.N. gratefully acknowledges support by the Deutsche Forschungsgemeinschaft (DFG, German Research Foundation) -- Project-ID 138713538 -- SFB 881 (''The Milky Way System'', subproject B8). R.P. and S.K. gratefully acknowledge funding from UKRI in the form of a Future Leaders Fellowship (grant no. MR/T022868/1. JS acknowledges support from NSF grant AST-1812856 and the Packard Foundation. This paper is based on observations obtained at the international Gemini Observatory, a program of NSF’s NOIRLab, which is managed by the Association of Universities for Research in Astronomy (AURA) under a cooperative agreement with the National Science Foundation. on behalf of the Gemini Observatory partnership: the National Science Foundation (United States), National Research Council (Canada), Agencia Nacional de Investigaci\'{o}n y Desarrollo (Chile), Ministerio de Ciencia, Tecnolog\'{i}a e Innovaci\'{o}n (Argentina), Minist\'{e}rio da Ci\^{e}ncia, Tecnologia, Inova\c{c}\~{o}es e Comunica\c{c}\~{o}es (Brazil), and Korea Astronomy and Space Science Institute (Republic of Korea).

\bibliography{refs.bib}

\end{document}